\documentclass[aps,prd
,preprint,tightenlines,
nofootinbib,showpacs]{revtex4}
\usepackage{amssymb,latexsym}
\usepackage{amsmath,amsbsy,bbm}
\usepackage{epsfig,bm}
\usepackage{graphicx,comment}
\DeclareMathOperator{\tr}{tr}
\DeclareMathOperator{\Erfc}{Erfc}

\begin{document}
\def\a{{\alpha}}
\def\b{{\beta}}
\def\d{{\delta}}
\def\D{{\Delta}}
\def\e{{\varepsilon}}
\def\g{{\gamma}}
\def\G{{\Gamma}}
\def\k{{\kappa}}
\def\l{{\lambda}}
\def\L{{\Lambda}}
\def\m{{\mu}}
\def\n{{\nu}}
\def\o{{\omega}}
\def\O{{\Omega}}
\def\S{{\Sigma}}
\def\s{{\sigma}}
\def\th{{\theta}}

\def\ol#1{{\overline{#1}}}

\def\Dslash{D\hskip-0.65em /}
\def\Dtslash{\tilde{D} \hskip-0.65em /}

\def\CPT{{$\chi$PT}}
\def\QCPT{{Q$\chi$PT}}
\def\PQCPT{{PQ$\chi$PT}}
\def\tr{\text{tr}}
\def\str{\text{str}}
\def\diag{\text{diag}}
\def\order{{\mathcal O}}

\def\meff{{m^2_{\text{eff}}}}

\def\Meff{{M_{\text{eff}}}}
\def\cF{{\mathcal F}}
\def\cS{{\mathcal S}}
\def\cC{{\mathcal C}}
\def\cE{{\mathcal E}}
\def\cB{{\mathcal B}}
\def\cT{{\mathcal T}}
\def\cQ{{\mathcal Q}}
\def\cL{{\mathcal L}}
\def\cO{{\mathcal O}}
\def\cA{{\mathcal A}}
\def\cV{{\mathcal V}}
\def\cR{{\mathcal R}}
\def\cH{{\mathcal H}}
\def\cW{{\mathcal W}}
\def\cM{{\mathcal M}}
\def\cD{{\mathcal D}}
\def\cN{{\mathcal N}}
\def\cP{{\mathcal P}}
\def\cK{{\mathcal K}}
\def\Qt{{\tilde{Q}}}
\def\Dt{{\tilde{D}}}
\def\St{{\tilde{\Sigma}}}
\def\cBt{{\tilde{\mathcal{B}}}}
\def\cDt{{\tilde{\mathcal{D}}}}
\def\cTt{{\tilde{\mathcal{T}}}}
\def\cMt{{\tilde{\mathcal{M}}}}
\def\At{{\tilde{A}}}
\def\cNt{{\tilde{\mathcal{N}}}}
\def\cOt{{\tilde{\mathcal{O}}}}
\def\cPt{{\tilde{\mathcal{P}}}}
\def\cI{{\mathcal{I}}}
\def\cJ{{\mathcal{J}}}

\def\eqref#1{{(\ref{#1})}}

\preprint{UMD-40762-401}
 
\title{External Momentum, Volume Effects, and the Nucleon Magnetic Moment}

\author{Brian C.~Tiburzi}
\email[]{bctiburz@umd.edu}
\affiliation{%
Maryland Center for Fundamental Physics, 
Department of Physics, 
University of Maryland, 
College Park,  
MD 20742-4111, 
USA
}

\date{\today}


\begin{abstract}
We analyze the determination of volume effects for 
correlation functions that depend on an external momentum.
As a specific example, we consider finite volume 
nucleon current correlators, and focus on
the nucleon magnetic moment. 
Because the multipole decomposition relies on $SO(3)$
rotational invariance, the structure of such finite volume corrections
is unrelated to infinite volume multipole form factors. 
One can deduce volume corrections to the magnetic 
moment only when a zero-mode photon coupling vanishes, 
as occurs at next-to-leading order
in heavy baryon chiral perturbation theory. 
To deduce such finite volume corrections, however, 
one must assume continuous momentum transfer.
In practice, volume corrections with momentum transfer
dependence are required to address the extraction of the 
magnetic moment, or other observables that arise in momentum
dependent correlation functions. 
Additionally we shed some light on a puzzle concerning  
differences in lattice form factor data at equal values 
of momentum transfer squared. 
\end{abstract}
\maketitle

\section{Introduction}

QCD in the non-perturbative regime
has proven notoriously difficult to understand
quantitatively. 
This non-Abelian gauge theory of gluons coupled to quarks 
is simple enough to write down, and even becomes 
perturbative at high energies. 
At low energies, 
however, the theory is strongly coupled
which results in the confinement of quarks and gluons
into hadrons.  
Accounting for the measured properties of hadrons is a challenge; 
making reliable predictions is even more difficult. 
After more than a quarter century of dedicated work, 
lattice gauge theory has made considerable
progress in addressing  strong interaction physics.
As a first principles
numerical technique,
lattice gauge theory can be used to 
determine low-energy hadronic
properties rigorously from QCD. 
For a comprehensive overview of 
lattice methods, see~\cite{DeGrand:2006zz}

Despite considerable success, lattice QCD 
calculations still suffer from a number of systematic
errors.  Due to computational restrictions, 
simulations cannot be carried out at the physical pion mass. 
Instead heavier pions must be used.
Additionally current lattice volumes are not significantly
larger than the typical hadronic length scale, which for most
observables is set by the pion Compton wavelength. 
To eliminate such systematic errors, it is desirable to have
an independent tool that predicts the pion mass, 
and lattice volume dependence of observables. 
Fortunately, effective field theory techniques
exist suited for this purpose.  Chiral perturbation 
theory, for example, provides a model 
independent and systematic tool to address
the pion mass and lattice volume dependence of
many low-energy hadronic properties.
Pioneering work on the volume dependence 
of the chiral condensate and pion observables 
appeared a while ago in series of papers%
~\cite{Gasser:1986vb,Gasser:1987ah,Leutwyler:1987ak,Gasser:1987zq}.
Additionally finite volume amplitudes have been analyzed 
in order to study unstable particles and multiparticle physics%
~\cite{Luscher:1990ux,Luscher:1991cf,Lellouch:2000pv}.
A variety of observables have since been treated 
in finite volume. Recent work has included treating heavy mesons and 
baryons~\cite{Khan:2003cu,Arndt:2004bg,Beane:2004tw,Beane:2004rf,Bernard:2007cm}, 
extending the validity of finite volume corrections for meson masses and 
decay constants~\cite{Colangelo:2004xr,Colangelo:2005gd,Colangelo:2006mp},
exploring different regimes in finite volume 
theories~\cite{Detmold:2004ap,Bedaque:2004dt,Smigielski:2007pe},
and investigating multiparticle systems in finite 
volume~\cite{Beane:2003da,Bedaque:2006yi,Sato:2007ms,Beane:2007qr}.

Further work at finite volume has included processes depending on 
an external momentum~%
\cite{Detmold:2005pt,Bunton:2006va,Chen:2006gg,Detmold:2006vu}.
Deducing volume corrections to electromagnetic observables, for example,
can be a subtle task~\cite{Hu:2007eb, Hu:2007ts}. 
A point that was specifically addressed in these latter works 
was the lacking connection of finite volume amplitudes
to low-energy multipole  expansions. 
Multipole expansions are inherent in the 
description of $SO(3)$ invariant physics. 
At finite volume such a description of low-energy
matrix elements ceases to be valid.
Utilizing effective field theory,  finite volume corrections to 
amplitudes can be computed, but not corrections to 
multipole moments and other observables, 
such as multipole polarizabilities. 
Said another way, an effective field theory only
allows one to match the calculation of correlation
functions. 
Thus chiral perturbation theory allows one to calculate 
the model-independent, long-distance behavior of 
finite volume correlation functions.
Connection to infinite volume observables,
which might be accessible through various 
different correlation functions, is not mandated.

In this work, we consider finite volume correlation functions 
depending on an external momentum. Specifically, we detail the finite volume 
modifications to nucleon current matrix 
elements, and analyze whether it is possible
to deduce volume corrections to the nucleon
magnetic moment. 
Based on symmetry considerations, we show that 
such a connection to infinite volume physics is not generally possible;
but, within the framework of heavy baryon 
chiral perturbation theory, the vanishing of 
a new finite volume coefficient at next-to-leading
order allows for the connection to be made. 
We are careful to expose, however, 
that this analysis relies on the assumption of 
continuous momentum transfer. 
On the lattice, the available momentum modes
are quantized, but there is a regime 
in which the assumption of continuous momentum is approximately valid. 
This regime, however, is beyond the reach of
current computing resources. 
Thus in practice, the volume effect must be
determined from the momentum transfer dependent
nucleon current matrix element.%
\footnote{We mention that background field methods alternatively
utilize two-point functions to deduce magnetic
moments from the shift
in particle energies linear in the magnetic field, 
see e.g.~\cite{Martinelli:1982cb,Bernard:1982yu,Lee:2005ds}. 
As we focus on three-point functions, 
our analysis has no direct relation to 
finite volume corrections in
background field calculations.}
Although we have specialized to current matrix elements, 
our analysis generalizes to any momentum-dependent 
correlation function at finite volume. 
Finally using our expressions for current matrix elements, 
we show that form factors calculated at equal values of 
momentum transfer squared (but at differing momentum transfer)
can differ due to volume effects. This potentially resolves a puzzle seen 
in lattice form factor data.

Our presentation has the following organization.
In Sect.~\ref{SPEA}, we write down the single particle
effective action for the nucleon coupled to 
an electromagnetic field. This is done for a
spatial torus. Here we show based on symmetry 
arguments that the electromagnetic
current is additively renormalized, and that 
there is an additional magnetic-like zero-mode
interaction. 
Next in Sect.~\ref{HB}, we calculate polarized 
nucleon electromagnetic current  matrix elements
in finite volume. Heavy baryon
chiral perturbation theory is utilized, and the relevant 
finite volume functions appearing in the calculation are listed in Appendix~\ref{FVF}. 
The results of this 
calculation can be used to determine the volume-dependent
coupling constants in the single nucleon effective action. 
In Sect.~\ref{MDR}, we investigate the conditions necessary
for the matching to be performed, and find they
require lattice sizes larger than currently available. 
In Appendix~\ref{B}, we carry out the momentum expansion
one non-trivial order further to discuss the magnetic radius 
at finite volume. 
The form factor difference puzzle is taken up in Sect.~\ref{puzzle}. 
Generic features of our findings are summarized in Sect.~\ref{summy}.

\section{Single Particle Effective Action} \label{SPEA}

We begin by considering the low-energy dynamics 
of a nucleon in an external electromagnetic field. 
We use a two component isospinor $N$ for the nucleon.
In infinite volume the nucleon can be described
by the Lagrangian
\begin{equation} \label{eq:EFT}
\cL 
=
\ol N 
\left(
i  v \cdot D 
- 
\frac{D_\perp^2}{2 M_N}
-
\frac{i \mu_N}{2 M_N}  
\left[ S_\mu, S_\nu \right]
F^{\mu \nu}
\right)
N
,\end{equation}
where 
$v^\mu$ 
is the heavy nucleon four-velocity,
$S^\mu$ is the covariant spin operator,  and 
$D_\perp^\mu = D^\mu - v^\mu (v \cdot D)$, 
see~\cite{Jenkins:1991jv}.
The coefficient of the kinetic term is exactly fixed by reparametrization 
invariance~\cite{Luke:1992cs}, but the form of the operator 
is unique only up to field redefinitions. 
The gauge covariant derivative appearing above is 
$D_\mu = \partial_\mu + i e Q_N A_\mu$, 
where 
$Q_N$ 
is the nucleon charge matrix. 
The coupling constant
$\mu_N$ 
is a diagonal matrix containing the proton 
and neutron magnetic moments. 
In writing the above Lagrangian, we have included all zeroth 
and first order terms in an expansion in photon frequency.
At higher order in the frequency expansion, there 
are terms with more powers of the 
field strength tensor, 
$F^{\mu \nu}$, and derivatives thereof. 
Because we work well below the pion production threshold, 
we have integrated out pion-nucleon interactions
to arrive at Eq.~\eqref{eq:EFT}. 
The low-energy constants in the matrix 
$\mu_N$ 
depend on the pion mass and other couplings
of the nucleon theory with pion interactions. 
That theory is chiral perturbation theory,
and we treat such dependence as implicit.

In finite volume, we can write down a theory 
analogous to Eq.~\eqref{eq:EFT}.  To be concrete, 
we assume that this theory is defined on a torus of length 
$L$ 
in each of the three spatial directions. This reflects the underlying 
lattice QCD action with quarks subject to periodic boundary conditions,
but ignoring any possible effect from the finite time direction. 
The general form of the single nucleon effective action 
can be constructed by writing down all possible operators
consistent with the symmetries of electromagnetism on a torus. 
The discrete symmetries 
$C$, 
$P$, 
and 
$T$ 
remain. Boost invariance and  
$SO(3)$ 
rotational invariance 
possessed by the Lagrangian in Eq.~\eqref{eq:EFT}, however,
are reduced to the cubic symmetry group, which is isomorphic to 
$S_4$. 
On compactified spaces,
gauge transformations are more constrained for the gauge field zero mode. 
A well known example of this occurs in
finite temperature gauge theory, see, e.g.~\cite{Zinn-Justin:1996cy}.
On the spatial torus we consider, 
the zero modes of the three-vector potential have 
periodicity constraints under gauge transformations. 
This restriction on gauge transformations leads to the 
ability to construct more gauge invariant operators than in 
infinite volume.

In~\cite{Hu:2007eb}, the spatial analog of 
the Polyakov line was used to write down gauge invariant 
operators involving the gauge field zero modes. 
Specifically employed was the Wilson line 
$\cW_i$
defined by
\begin{equation} \label{eq:Wline}
\cW_i =  \exp  \left( \frac{ i e}{3}  \oint dx_i \, A_i \right)
,\end{equation}
where the line integral starts at some point 
$x_\mu$
and runs to 
$x_\mu + L \hat{x}_i$. 
There is no implied sum over 
$i$ 
on the right-hand side of Eq.~\eqref{eq:Wline}.
Because the gauge potential
is periodic,\footnote{%
In lattice simulations that probe electromagnetic observables,
either a background electromagnetic field is gauged into the
action, or current matrix elements are calculated. In the latter
case, the current operator 
$J_\mu (x) = \ol \psi(x) \gamma_\mu \psi(x)$
is periodic. Hence in an effective theory for such calculations,
one must take 
$A_\mu$
to be periodic otherwise the effective action is not single valued. 
Our discussion, however, does not apply to the background field method 
which requires a separate treatment altogether.
} 
cycling the compact dimension must produce 
a gauge invariant object. The factor of
$e/3$ 
reflects that
the quark charges are quantized in such units. 
Under a gauge transformation, we have 
$A^\mu \to A^\mu + \partial^\mu \alpha$. 
In order for the Wilson line in  Eq.~\eqref{eq:Wline}
to be gauge invariant, we must have
\begin{equation}
\alpha(x_i = L) = \alpha  (x_i = 0) + \frac{6 \pi}{e} n_i
,\end{equation}
where 
$n_i$ 
is an integer. As this condition must hold
in each spatial direction, we find
\begin{equation}
\alpha(x) = \frac{6 \pi}{e L} \bm{n} \cdot \bm{x} + \ol \alpha(x) 
,\end{equation}
where 
$\ol \a(x)$ 
is a periodic function of the 
$x_i$. 
The linear term in the gauge function 
$\a(x)$ 
gives rise to a quantized shift of the gauge field zero mode.

To write an effective theory of the nucleon and photons
in finite volume, it is easiest to use linear combinations of 
Wilson lines that are Hermitian and have definite $C$, $P$, 
and $T$ transformations. 
These are the operators $\cW_i^{(+)}$, and $\cW_i^{(-)}$
given by
\begin{eqnarray}
\cW_i^{(+)} &=&  \frac{1}{2} \left( \cW_i + \cW_i^\dagger \right),
\\
\cW_i^{(-)} &=& \frac{1}{2 i} \left( \cW_i - \cW_i^\dagger \right)
.\end{eqnarray}
It is easy to show that these two operators are not independent,
and we choose to work with the 
$\cW_i^{(-)}$ 
because they have the same
$C$, 
$P$,
 and 
 $T$ symmetry transformations as the 
 $A_i$. 
Constructing the most general gauge invariant Lagrangian
on a torus is arbitrarily complicated. As our interest is with the nucleon
magnetic moment, we restrict our attention to operators having only
one insertion of 
$\cW_i^{(-)}$. 
Multiple insertions of
$\cW_i^{(-)}$ 
lead to photon-nucleon
couplings with more than one photon. Furthermore we restrict our attention
to operators with at most one derivative.%
\footnote{Terms with further derivatives are considered in Appendix~\ref{B}. 
There we write down the relevant terms
for the volume corrections to the magnetic radius.} 
We will investigate the conditions
that justify such a photon frequency expansion in Sect.~\ref{MDR}.

We now build the single particle effective action for nucleons 
and photons on a torus. 
To write this theory, we abandon the 
covariant notation employed in Eq.~\eqref{eq:EFT} above, 
because the torus has only 
$S_4$ 
invariance. Including all single 
$\cW^{(-)}_i$ operators
with at most one derivative, we have the Lagrangian
\begin{equation} \label{eq:EFTL}
\cL
=
\ol N
\left[
i D_0
+
\frac{\bm{D}^2}{2 M_N}
-
\frac{C_1(L) }{2 M_N} 
i \overset{\leftrightarrow}{\bm{D}} 
\cdot 
\bm{\cW}^{(-)}
+ 
\frac{\mu_N(L)} {2 M_N} \,
\bm{\sigma} \cdot \bm{B}
+
\frac{ C_2(L)}{ 2 M_N} \,
\bm{\sigma} \cdot \left( \bm{\nabla} \times \bm{\cW}^{(-)} \right)
\right]
N
,\end{equation}
where 
$i \overset{\leftrightarrow}{\bm{D}} 
= 
i 
\left(
\overset{\leftarrow}{\bm{D}} 
-
\overset{\rightarrow}{\bm{D}}
\right)
$ defines an Hermitian operator. 
Notice that 
$\nabla_i  \cW^{(-)}_i = 0$.
In writing Eq.~\eqref{eq:EFTL}, we have explicitly indicated the 
dependence on the spatial length 
$L$. 
The coupling constants
above run with the infrared cutoff,
$1/L$, 
and must be determined
by matching calculations in the microscopic theory. 
For example, the magnetic
moment operator is accompanied by the finite
volume coefficient 
$\mu_N(L)$ 
which is given by
\begin{equation}
\mu_N(L) = \mu_N + \delta \mu_N(L)
,\end{equation}
where the term 
$\delta \mu_N(L)$ 
is the finite volume effect
that can be determined using chiral perturbation theory.
Running the infrared cutoff to zero produces the infinite 
volume magnetic moment, 
$\underset{L \to \infty}{\lim} \mu_N(L)  =   \mu_N$.
Compared to Eq.~\eqref{eq:EFT}, there 
are two new operators allowed by symmetry.
These operators contain single photon couplings as well
as a tower of cubic invariant multi-photon couplings.
There are, however, further multi-photon operators that we have
not written in Eq.~\eqref{eq:EFTL}. Such operators involve 
multiple insertions of 
$\cW^{(-)}_i$.
The new coupling constants 
$C_1(L)$ 
and 
$C_2(L)$ appearing in 
Eq.~\eqref{eq:EFTL} both 
must run to zero when the infinite volume limit is taken.
As we work below any multi-particle thresholds, these
$L$-dependent couplings run to zero exponentially fast in 
asymptotically large volumes~\cite{Luscher:1991cf}.
Notice that expanding the Lagrangian in Eq.~\eqref{eq:EFTL} to linear order 
in the gauge field, we have an accidental 
$SO(3)$ 
invariance.

The effective theory in Eq.~\eqref{eq:EFTL} can be used to 
calculate single photon-nucleon processes. For example, 
for an unpolarized nucleon at rest, we have the current matrix element
\begin{equation}
\langle 
N(\bm{0}) 
| J^\mu | 
N(\bm{0})
\rangle
= 
Q_N \, e \,  g^{\mu 0}
,\end{equation}
which produces the total charge. 
In infinite volume,  we can boost this result 
to an inertial frame where the nucleon
moves with momentum $\bm{p} = M_N \, \bm{v}$. 
By Galilean invariance, we expect the current 
to be $\bm{j} = Q_N e \bm{v}$. 
Using Eq.~\eqref{eq:EFTL} to calculate the current
in this frame, however, yields
\begin{equation} \label{eq:screen}
\langle 
N(\bm{p}) 
| \bm{J} | 
N(\bm{p})
\rangle
= 
[ Q_N - \cQ_N(L) ] e \bm{v}
,\end{equation}
where $\cQ_N(L) =  L \, C_1(L) / 3$. 
The electromagnetic current has been additively renormalized by
the operator with coefficient $C_1 (L)$. 
The theory described by Eq.~\eqref{eq:EFTL} 
is not Galilean invariant. 
This is the non-relativistic analog of  Lorentz symmetry violation
on a torus that allows current renormalization described in~\cite{Hu:2007eb}.
The sign in Eq.~\eqref{eq:EFTL} [and consequently that in Eq.~\eqref{eq:screen}]
anticipates that the infinite volume current is screened in finite volume.

We can further apply the effective Lagrangian in Eq.~\eqref{eq:EFTL} 
to polarized matrix elements. For a momentum transfer of 
$\bm{q} = \frac{2 \pi}{L} \bm{n}$, 
we find
\begin{equation}
\langle
N(\bm{q})  \downarrow
| J_k | 
N(\bm{0})  \downarrow
\rangle
- 
\langle
N(\bm{q})  \uparrow
| J_k |
N(\bm{0})  \uparrow
\rangle 
=
\frac{i  \varepsilon_{3 j k}  q_j }
{M_N}
\left[
\mu_N (L) 
+
\delta_{n_k,0} \,
\ol \mu_N (L)
\right] 
\label{eq:magmom}
,\end{equation}
which is sensitive to the finite volume magnetic moment 
and the additional zero-mode coupling 
$\ol \mu_N(L)$, 
which is given by 
$\ol \mu_N(L) = e L \, C_2(L) / 3$. 
To determine the volume dependent couplings, 
we must match finite volume chiral perturbation theory
calculations onto the single nucleon effective action in 
Eq.~\eqref{eq:EFTL}. 
We now turn to these calculations.

\section{Heavy Baryon Calculation} \label{HB}

\subsection{Chiral Perturbation Theory}

To determine volume corrections to photon-nucleon 
couplings, we utilize heavy baryon chiral perturbation theory. 
The virtual pion cloud of the nucleon deforms in finite volume 
because pions are the lightest hadrons, and easily propagate to the boundary. 
Their small masses arise because pions are the pseudo-Goldstone bosons arising 
from spontaneous chiral symmetry breaking. For the case
of two flavors, the symmetry breaking pattern is
from 
$SU(2)_L \otimes SU(2)_R \to SU(2)_V$, 
and the Goldstone manifold is parametrized by the coset field 
$\S \equiv \xi^2 =  \exp ( 2 i \phi / f)$, 
where the 
$SU(2)$ 
matrix 
$\phi$ 
contains the pions 
\begin{equation}
\phi
= 
\begin{pmatrix}
\frac{1}{\sqrt{2}} \pi^0 
&&
\pi^+
\\
\pi^- 
&&
- \frac{1}{\sqrt{2}} \pi^0
\end{pmatrix}
.\end{equation}
At leading order, 
$\cO(\varepsilon^2)$, 
in a small momentum, 
$p$, and pion mass,
$m_\pi$ 
expansion, 
where $\varepsilon \sim p \sim m_\pi$, 
the theory of pions is described by the Lagrangian
\begin{equation} \label{eq:ChPT}
\cL 
= 
\frac{f^2}{8} 
\left[
\tr ( D_\mu \S D^\mu )
+ 
m_\pi^2 \,
\tr ( \S + \S^\dagger )
\right]
.\end{equation}
Here we work in the isospin limit, and at tree level 
$m_\pi$ 
is the physical pion mass. The dimensionful parameter
$f$ 
is the pion decay constant, 
$f = 132 \, \texttt{MeV}$. 
Electromagnetism
has been gauged into Eq.~\eqref{eq:ChPT} via the covariant derivative
$D^\mu = \partial^\mu + i e A_\mu [ Q, \, \, \,]$, where 
$Q$ 
is the quark electric charge matrix
\begin{equation}
Q 
= 
\diag \left( \frac{2}{3}, - \frac{1}{3} \right)
.\end{equation}

To include baryons we use the isospinor $N$ from above,
\begin{equation}
N 
= 
\begin{pmatrix}
p \\
n
\end{pmatrix}
,\end{equation}
as well as the heavy baryon formulation which allows us
a consistent expansion in powers of residual momentum, $k_\mu \sim \varepsilon$, 
where we decompose an arbitrary nucleon momentum $P$ into 
$P_\mu = M_N\,  v_\mu + k_\mu$. 
To leading order, $\cO(\varepsilon)$, the heavy baryon chiral 
Lagrangian is
\begin{equation} \label{eq:B}
\cL 
= 
\ol N
\left(
i v \cdot D 
+ 
2 g_A 
S \cdot \cA
\right) N
,\end{equation}
where $\cA_\mu$ and $\cV_\mu$ are axial-vector and vector
pion fields defined by
\begin{eqnarray}
\cA_\mu &=& \frac{i}{2} 
\left(
\xi \partial_\mu \xi^\dagger
- 
\xi^\dagger \partial_\mu \xi 
\right),
\\
\cV_\mu &=& \frac{1}{2}
\left(
\xi \partial_\mu \xi^\dagger
+ 
\xi^\dagger \partial_\mu \xi 
\right)
,\end{eqnarray}
the latter appears in the chirally and electromagnetically gauge
covariant derivative $D_\mu$, which acts on the nucleon
isospinor as
\begin{equation}
D_\mu N 
= 
\left(
\partial_\mu + \cV_\mu + i e Q_N  A_\mu
\right)
N
.\end{equation}
The nucleon charge matrix $Q_N$ is given by $Q_N = Q + \tr \, Q$.

The nearest baryon resonances in the spectrum of QCD are the deltas.
Phenomenologically we know the nucleon-delta
axial couplings are not small. Furthermore the
nucleon-delta mass splitting $\Delta \sim 300 \, \texttt{MeV}$
is not much greater than the pion mass. In fact 
on current lattices, the lightest lattice pion masses 
are the same size as $\Delta$. Hence for a consistent
power counting, we must treat $\Delta \sim \varepsilon$, 
and retain deltas in the theory as explicit degrees of 
freedom.  The deltas are described by the flavor tensor field
$T_\mu$, which is heavy baryon Rarita-Schwinger spinor.
The leading order, $\cO(\e)$, Lagrangian for the deltas
and their interactions with nucleons is given by
\begin{equation} \label{eq:T}
\cL 
=
- \ol T {}^\mu
\left(
 i v \cdot D - 
\Delta 
\right) 
T_\mu
+ 
g_{\Delta N} 
\left(
\ol T {}^\mu \cA_\mu N 
+
\ol N \cA^\mu T_\mu
\right)
+ 
2 g_{\Delta \Delta}
\ol T {}^\mu S \cdot \cA T_\mu
.\end{equation}

\subsection{Magnetic Form Factor}

Having spelled out the interactions of nucleons with 
pions and deltas, we detail the calculation of the magnetic
form factor in infinite volume. Consider
the polarized nucleon matrix elements
that define the Pauli form factor
\begin{equation}
\langle
N(\bm{q})  \downarrow
| J_k | 
N(\bm{0})  \downarrow
\rangle
- 
\langle
N(\bm{q})  \uparrow
| J_k |
N(\bm{0})  \uparrow
\rangle 
=
\frac{i  q_j \varepsilon_{3 j k}}
{M_N}
F_2(q^2)
\label{eq:spinpol}
.\end{equation}
The magnetic moment is given by 
the form factor at zero momentum transfer
\begin{equation}
\mu_N  = F_2(0)
.\end{equation}
In the heavy baryon theory, 
contributions to 
$F_2 (q^2)$
arise from local photon-nucleon 
interactions, 
as well  as from virtual pion loop 
contributions generated by 
leading-order Lagrangian. 
The leading contribution
to magnetic moments come from 
the local interactions at
$\cO(\e^0)$. These are contained 
in the current operator
\begin{equation}
J^\mu
= 
-
\frac{i \partial_\nu}{M_N}
\left(
\mu_0 
\ol N  [S^\mu, S^\nu ] N
+
\mu_I
\ol N [S^\mu, S^\nu ] \tau^3 N
\right)
,\end{equation}
where $\tau^3$ is an isospin matrix. 
The two terms above contribute to the isoscalar
and isovector moments, respectively. 
For simplicity we shall address the isovector magnetic
moment and isovector Pauli form factor. These are defined 
simply as proton minus neutron differences in these respective 
quantities.

\begin{figure}
\epsfig{file=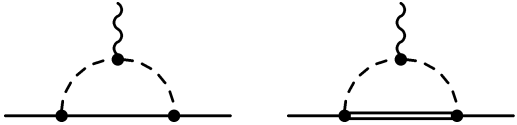}
\caption{Loop diagrams contributing to the nucleon magnetic moments at $\cO(\e)$. 
The photon is depicted as a wiggly line, mesons are denoted by a dashed line,  
and a thin solid line denotes a nucleon, while the double line denotes a delta. 
}
\label{f:magmom}
\end{figure}

Higher order corrections to the Pauli form factor
come from loop diagrams generated from the leading
order Lagrangian in Eqs.~\eqref{eq:B}, and~\eqref{eq:T}. 
Specifically the one loop graphs for the Pauli 
form factor are depicted in Figure~\ref{f:magmom}.
These graphs make $\cO(\e)$ contributions to magnetic
moments. 
Evaluation of these diagrams, combined with the tree-level result
yields the isovector Pauli form factor $F_2^{(I)}(q^2)$~%
\cite{Jenkins:1992pi,Bernard:1992qa,Bernard:1998gv},
\begin{equation} \label{eq:Iso}
F_2^{(I)}(q^2)
=
2 \mu_I
- 
\frac{g_A^2 M_N}{2 \pi f^2}
\int_0^1 dx \, m_\pi P_\pi (x, q^2)
- 
\frac{g_{\Delta N}^2 M_N}{9 \pi^2 f^2}
\int_0^1 dx \, F[ m_\pi P_\pi (x,q^2), \Delta ]
,\end{equation}
where 
\begin{equation} \label{eq:Ppi}
P_\pi (x, q^2) = \sqrt{1 - x (1-x) \frac{q^2}{m_\pi^2}}
,\end{equation}
encodes the momentum transfer dependence, 
and the non-analytic function 
$F(m, \delta)$ 
is given by
\begin{equation}
F(m, \delta) 
= 
- \delta 
\log
\left(
\frac{m^2}{4 \delta^2}
\right)
+ 
\sqrt{\delta^2 - m^2}
\log 
\left(
\frac{\delta - \sqrt{\delta^2 - m^2} + i \epsilon}{\delta + \sqrt{\delta^2 - m^2} + i \epsilon}
\right)
.\end{equation}
In order to arrive at this answer we have renormalized the tree-level
coupling 
$\mu_I$, 
so that at one-loop order the 
$\mu_I$ 
in Eq.~\eqref{eq:Iso}
is the chiral limit value.

\subsection{Finite Volume Form Factor}

We now consider the finite volume modifications
to the isovector Pauli form factor.  We delay 
the determination of volume corrections to the 
magnetic moment until Sect.~\ref{MDR}. 
To calculate volume corrections, we use the 
same setup as described in Sect.~\ref{SPEA}, 
namely we take a finite space of volume 
$L^3$
with an infinite time extent. The pions, nucleons, 
and deltas, moreover, have periodic boundary conditions
which stem from the periodicity of quarks in the lattice
action, and the fact that hadron fields are point-like 
objects in the effective theory.

To calculate observables in the finite volume theory
we use the Lagrangian in Eqs.~\eqref{eq:ChPT}, \eqref{eq:B}, 
and \eqref{eq:T}. The action, however, is given by 
the volume intergral of the Lagrangian with periodic 
boundary conditions on all fields enforced. Thus 
the same Feynman diagrams are generated in the 
finite volume theory as in infinite volume. 
The difference is the spatial momentum
quantization of internal and external lines. 
We must further choose 
$m_\pi L \gg 1$ so that the zero mode of the pion field  
does not become strongly coupled~\cite{Gasser:1986vb,Gasser:1987ah}. 
With this assumption on the lattice size 
$L$,
the 
$\varepsilon$
power counting we have
employed in infinite volume carries over to finite volume~\cite{Gasser:1987zq}.

The volume correction to the nucleon current matrix element, 
$\delta \langle N  | J^\mu | N  \rangle_L$, 
can be determined by a trivial matching condition, namely
\begin{equation} \label{eq:cartoon}
\langle N  | J^\mu | N  \rangle_L
= 
\langle N  | J^\mu | N  \rangle_\infty
+ 
\d \langle N  | J^\mu | N  \rangle_L
,\end{equation}
where 
$\langle N  | J^\mu | N  \rangle_L$
is the finite volume current matrix element, and
$\langle N  | J^\mu | N  \rangle_\infty$ 
is the infinite volume result, e.g.~the isovector
magnetic contribution shown in Eq.~\eqref{eq:Iso}. 
We have temporarily suppressed the momentum transfer argument
and denote $L$-dependence with subscripts rather than parenthetically. 
Because the finite and infinite volume theories share exactly the same
ultraviolet divergences, the matching condition above ensures
that $\d \langle N  | J^\mu | N  \rangle_L$ is indeed the infrared effect.

Carrying out the matching in Eq.~\eqref{eq:cartoon} 
on the spatial isovector current matrix element, we find
\begin{eqnarray} 
\d \langle N(\bm{q}) | J^i_{\text{Iso}} | N (\bm{0}) \rangle_L 
&=&
- \frac{3}{f^2} 
\left[
g_A^2 
f^{ij} (q^2, \bm{q}, 0)
+ 
\frac{2}{9}
g_{\D N}^2 
f^{ij} (q^2, \bm{q}, \D)
\right]
\, 
\ol u  \left[ S^j, \bm{S} \cdot \bm{q}  \right] u
\label{eq:Pauli}
,\end{eqnarray}
where we have defined the function
\begin{equation}
f^{ij} (q^2, \bm{q}, \d )
= 
\int_0^1 dx 
\left\{
2 I^{ij}_{5/2} [x \bm{q}, m_\pi P_\pi(x,q^2), \d]
+ 
q^i I^j_{5/2} [x \bm{q}, m_\pi P_\pi(x,q^2), \d]
\right\} 
\label{eq:weird}
,\end{equation}
in terms of  $I^{i}_\b(\bm{\th}, m, \d)$, and $I^{ij}_\b(\bm{\th},m,\d)$
which are finite difference functions appearing in the Appendix. 
The $u$, and $\ol u$ appearing in Eq.~\eqref{eq:Pauli} are Pauli spinors, 
and $P_\pi(x, q^2)$ is given in Eq.~\eqref{eq:Ppi}.

From Eq.~\eqref{eq:Pauli}, we cannot identify 
finite volume corrections to the Pauli form factor. 
Indeed the form factor decomposition of 
nucleon current matrix elements is tied to
Lorentz and gauge invariance.  
The former does not hold on a torus, while 
the latter has a completely different form. 
The integrand of Eq.~\eqref{eq:weird} is clearly an 
even function of each component of $\bm{q}$.
More generally it is a cubic invariant function of 
the spatial momentum, and so depends on $\bm{q}^2$, 
as well as $\sum_i q_i q_i q_i q_i$, and a tower of other 
cubic invariant combinations. As there is no simple way 
to denote this, we have chosen simply to write $f^{ij} = f^{ij}(q^2, \bm{q}, \d)$. 
This function should be considered as a finite 
volume generalization of a form factor.

With Eq.~\eqref{eq:Pauli}, we have determined
volume corrections to nucleon current matrix elements. 
This result can be directly utilized to remove volume
dependence from lattice calculations of the nucleon 
current provided one is in the range of applicability
of heavy baryon chiral perturbation theory. 
The expression can be simplified, moreover, depending 
on the actual lattice kinematics used in measuring the 
three-point function on the lattice.  For example, to be sensitive 
to the magnetic coupling in Eq.~\eqref{eq:magmom},
the momentum transfer $\bm{q}$ must have at least 
one component transverse to the plane of spin polarization. 
We shall thus assume that $\bm{q}$ has no component
parallel to the spin polarization. Additionally from Eq.~\eqref{eq:magmom}, 
we see that the spatial component of the current measured
must be transverse to the spin polarization. 
We can thus take $\bm{q}$ and $\bm{J}$ to be
mutually orthogonal in the plane transverse to the polarization. 
This choice is sensitive to the Pauli form factor and simplifies
the volume dependence. Because $\bm{q}$ is orthogonal to 
$\bm{J}$, the second term in the integrand of $f^{ij}(q^2, \bm{q}, \d)$ 
is zero since $i$ labels the index of the current in Eq.~\eqref{eq:Pauli}.

For the sake of concreteness take $\bm{q} = (q,0,0)$ and consider
the $y$-component of the current. Then we have the volume effect
that we loosely denote by $\d F_2^{(I)} (q, L)$ for lack of a better symbol, 
\begin{eqnarray}
\d F_2^{(I)} (q, L)
&\equiv&
\frac{M_N}{i q}
\Big[ 
\d \langle N(\bm{q}) \downarrow | \hat{\bm{y}} \cdot \bm{J}_{\text{Iso}} | N (\bm{0}) \downarrow \rangle_L 
- 
\d \langle N(\bm{q}) \uparrow | \hat{\bm{y}} \cdot \bm{J}_{\text{Iso}} | N (\bm{0}) \uparrow \rangle_L 
\Big]
\notag \\
&=& - 
\frac{6 M_N}{f^2}
\int_0^1 dx 
\left\{
g_A^2 
I^{22}_{5/2} 
[x q \hat{\bm{x}}, m_\pi P_\pi (x, q^2), 0]
+ 
\frac{2}{9}
g_{\D N}^2
I^{22}_{5/2} 
[x q \hat{\bm{x}}, m_\pi P_\pi (x, q^2), \D]
\right\}.
\notag \\
\label{eq:start}
\end{eqnarray}
If other kinematics are chosen, 
one must return to Eq.~\eqref{eq:Pauli}
and evaluate accordingly.

\section{Matching and Discussion of Results} \label{MDR}

Having deduced the finite volume modification to 
nucleon current matrix element, we can now 
analyze the volume dependence and make contact
with the single particle action in Eq.~\eqref{eq:EFTL}. 
Connection to the single particle action with spin-dependent
higher derivative terms is carried out in Appendix~\ref{B}.

\subsection{Matching}
To perform the matching, 
we return to the volume corrections 
under the general kinematics,  Eq.~\eqref{eq:Pauli},
and series expand in the momentum transfer. 
To find the leading term, we need to evaluate the
finite volume form factor 
$f^{ij}(q^2, \bm{q}, \d)$
at zero momentum, 
\begin{equation}
f^{ij} (0,\bm{0},\d)
=
\frac{2}{3}
\delta^{ij}
J_{5/2} (m_\pi, \d)
,\end{equation}
where $J_\b(m,\d)$ is given in Appendix~\ref{FVF}. 
After some algebra, the leading term in the series 
expansion of the matrix element can be written as 
\begin{eqnarray} 
\d \langle N(\bm{q}) | J^i_{\text{Iso}} | N (\bm{0}) \rangle_L 
&=&
\frac{1}{M_N} 
\ol u  \left[ S^i, \bm{S} \cdot \bm{q}  \right] u
\, \,
\d F^{(I)}_2 (0,L)
+ 
\cO(q^3)
\label{eq:PauliExp}
,\end{eqnarray}
where 
$F_2^{(I)}(0,L)$ 
is identical to the function in Eq.~\eqref{eq:start}
evaluated at 
$q=0$.
Comparing with Eq.~\eqref{eq:EFTL}, we thus find
\begin{eqnarray}
\mu_N^{(I)} (L) &=&  F_2^{(I)} (0) + \d F_2^{(I)} (0,L)
\\
C_2(L) &=& 0
.\end{eqnarray}
The term 
$\d F_2^{(I)} (0,L)$ 
is identified as the finite volume correction to the isovector magnetic moment,
$\d \mu_N^{(I)}(L)$, 
and (after algebra) is identical to that determined in~\cite{Beane:2004tw}. 
Furthermore the additional magnetic-like zero mode interaction in Eq.~\eqref{eq:EFTL}
vanishes to this order. This is true of both the isovector and isoscalar combinations. 
The remaining coefficient in Eq.~\eqref{eq:EFTL}, 
$C_1(L)$, 
must be determined
by computing the charge form factor in finite volume. 
The first non-vanishing contributions are from recoil order 
($\sim 1/M_N$) 
terms in they heavy baryon theory.

The finite volume correction to the current matrix element in Eq.~\eqref{eq:Pauli}
is determined at some general value $\bm{q} = 2 \pi \bm{n} / L$. 
We must investigate the conditions under which the series expansion in 
\emph{quantized} $\bm{q}$ leading to Eq.~\eqref{eq:PauliExp} is justified. 
For simplicity of notation, 
consider the momentum to lie entirely in one direction. 
A derivative expansion of the matrix element is justified as follows. 
For the $n$-th mode, $q_n = 2 \pi n / L$, and for large enough $n$, 
the relative difference $\d q$, 
\begin{equation}
\d q = \frac{q_{n+1} - q_{n}}{q_n} = \frac{1}{n} 
,\end{equation}
approaches zero. 
To rigorously series expand, we must take differences
of the nucleon current matrix element between adjacent 
modes for large mode number. This is the procedure implicitly 
needed to arrive at Eq.~\eqref{eq:PauliExp}.  Such a procedure
is lacking when one is restricted to just the lowest available
momentum mode, $q = 2 \pi / L$, see~\cite{Hu:2007ts}. We shall
investigate this below numerically.

We now estimate when Eq.~\eqref{eq:PauliExp}
gives a reasonable approximation to the finite volume effect.
For the effective theory to give a reasonable description
of the magnetic moment as opposed to the form factor,
we must have the higher order terms,
$\cO (q^3)$, under control. For $\sim 20 \%$ error in keeping 
the linear term in Eq.~\eqref{eq:PauliExp}, we require
\begin{equation}
\frac{q^2}{4 m_\pi^2} 
= 
\frac{\pi^2 n^2}{m_\pi^2 L^2}
\lesssim
\frac{1}{5}
.\end{equation}
To have a reasonable approximation to the derivative, say  
$\sim 20 \%$, 
we must further require
\begin{equation}
\d q  = \frac{1}{n} \lesssim \frac{1}{5}
.\end{equation}
Combining these two restrictions, we find
\begin{equation} \label{eq:est}
m_\pi L \gtrsim \sqrt{5} \pi n \gtrsim 5 \sqrt{5} \pi \approx 35 
.\end{equation}
Strictly speaking, this restriction must be met to determine the volume effect
to $\sim 20 \%$ using Eq.~\eqref{eq:PauliExp} 
(or using the equivalent expression in~\cite{Beane:2004tw})
for a lattice determination of the current matrix element using
the $n$-th mode.
The volume must be large so that there is a separation of scales:
the mode number must be small enough for the effective theory
to be applicable, and the mode number must be large enough 
to allow an approximation to a continuous derivative.
For volumes this large, however, the finite volume effect can be safely 
neglected altogether.

The actual
lattice determination of the nucleon 
magnetic moment complicates straightforward
application of the finite volume result, $\d \mu_N^{(I)}(L)$, derived above.
Current lattices are obviously too small to allow for the 
procedure outlined above to extract the moment. 
Instead one is left with quantized modes that are not
close enough together so that mode differences approximate derivatives.
Furthermore the values of the lowest non-vanishing momentum 
transfers are not in the range of applicability of 
the effective theory, which requires $q^2 / m_\pi^2 \sim 1$. 
Typically some modeling of the $q^2$-dependence of the matrix
element is done to extract the moment. 
It is unclear quantitatively what the 
volume correction  to such model-dependent extrapolations
should be. In order to shed some light on the subject, 
we analyze the finite volume correction to the current
matrix elements,
and give a qualitative picture for why $\d F_2^{(I)}(0,L)$ 
does not provide an estimate of volume effects relevant 
to typical lattice extractions of the magnetic moment.

\subsection{Discussion}
\begin{figure}
\epsfig{file=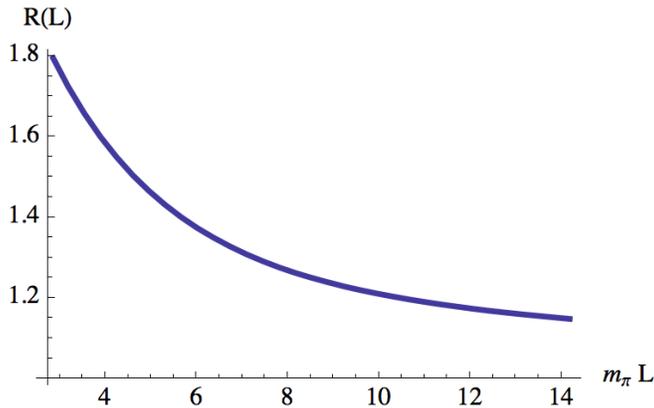,angle=270,width=9cm}
\caption{Plot of the ratio $R(L)$ in Eq.~\eqref{eq:R} as a function of $m_\pi L$. 
}
\label{f:R}
\end{figure}
Let us assume that the lattice practitioner 
has determined the isovector current utilizing 
the smallest non-vanishing momentum 
mode, $q = 2 \pi / L$. 
Consider the magnetic moment correction
as a correction to the finite volume current
rather than just narrowly to the magnetic moment.%
\footnote{ 
We shall often refer to current matrix elements and form
factors interchangeably. In the present case, the 
$F_2(q^2)$ form factor is just the current matrix element
scaled by $q = 2 \pi / L$. The same is true of the volume effects
due to the simplifying kinematics. 
}
For large enough box size, the smallest momentum mode
approaches zero, and so we expect $\d F_2^{(I)}(0,L)$ 
to give a good description of the volume dependence of
the finite volume form factor evaluated at the lowest mode. 
We can address this numerically
by comparing $\d F_2^{(I)}(0,L)$ to the momentum 
transfer dependent volume effect. For ease let us assume the 
lattice kinematics are such that we arrive at Eq.~\eqref{eq:start}
in finite volume. 
We plot the ratio $R(L)$ defined by
\begin{equation} \label{eq:R}
R(L) 
=
\frac{
\d F_2^{(I)}(0,L)
}
{ 
\d F_2^{(I)}(q,L) 
}
,\end{equation}
as a function of the box size $m_\pi L$. This is done in Figure~\ref{f:R}.
Here $q$ is fixed at the first non-vanishing momentum, 
and we use the values $g_A = 1.25$, $| g_{\D N} | = 1.5$,  
$M_N = 0.94 \, \texttt{GeV}$, and $m_\pi / \Delta = 0.48$. 
From the figure, we see that  $\d F_2^{(I)}(0,L)$ gives a 
rough estimate of the volume effect for the matrix element
under the simplifying kinematical choices. 
The agreement improves as the box size $L$ increases. 
The ratio $R(L)$ tends to one asymptotically but only very slowly.
Numerically one can see this asymptotic approach set in at $m_\pi L \approx 90$.  
As explained in~\cite{Hu:2007ts}, the current matrix element
depends on two combinations involving the momentum transfer:
$q^2 / m_\pi^2$, and  $\bm{q}^2 L^2$.  For the lowest non-vanishing
mode, the latter combination is constant and a series expansion
is poorly convergent.

\begin{figure}
\epsfig{file=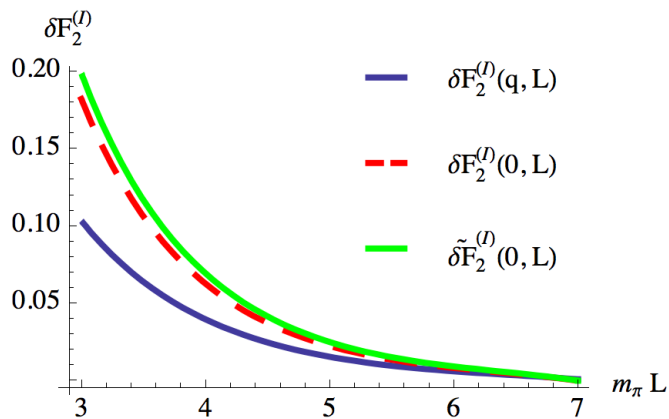,angle=270,width=9cm}
\caption{Plot of the finite volume modifications at $\bm{q} = \frac{2 \pi}{L} (1,0,0)$.
The finite volume matrix element 
$\d  F_2^{(I)} (q, L)$, 
the naive guess 
$\d  F_2^{(I)} (0, L) $, 
and the approximation
$\d \tilde F_2^{(I)} (0, L) $ 
are plotted  as a function of 
$m_\pi L$. 
}
\label{f:compare}
\end{figure}

An expansion in $q^2 / m_\pi^2$ for large $L$ is well behaved, and 
we can obtain a different approximation to $\d F_2^{(I)}(q,L)$
in the asymptotic regime. It is best to illustrate the issue by 
considering a similar but simpler example. Notice that for large $m_\pi L$, 
\begin{eqnarray} \label{eq:dipole}
P_\pi (x, q^2) \overset{L \to \infty} \longrightarrow 1
.\end{eqnarray}
Thus a typical finite volume contribution of the form
\begin{eqnarray}
\int_0^1 dx \, I_\b \left(   \frac{2 \pi n x}{L} \hat{\bm{x}}   ,  m_\pi P_\pi (x, q^2), \D \right)
\longrightarrow
\int_0^1 dx  \, I_\b \left(   \frac{2 \pi n x}{L} \hat{\bm{x}}   ,  m_\pi, \D \right)
,\end{eqnarray}
where we have specified the $n$-th mode in order to track the problematic
term. In this limit, the $x$-integral can now be performed because
the only $x$-dependence enters in one of the elliptic-theta functions $\vartheta_3$, 
namely as
\begin{eqnarray}
\int_0^1 dx \, \vartheta_3 ( \pi n x, e^{-\tau}) = 1
.\end{eqnarray}
Thus to take the large $L$ limit of $I_\b$, we merely drop the 
elliptic-theta function in the direction corresponding to the momentum 
transfer. The resultant volume correction is effectively two dimensional. 
This procedure is only valid when the momentum transfer 
is aligned with one of the spatial axes. The generally oriented case is considerably
more  complicated but the $x$-integration can still be performed.
Expanding in $q^2 / m_\pi^2$ but not in $\bm{q}^2 L^2$, we arrive at
\begin{equation}
\d \tilde F_2^{(I)} (0, L) 
= 
 - 
\frac{6 M_N}{f^2}
\int_0^1 dx 
\left\{
g_A^2 
I^{22}_{5/2} 
[x q \hat{\bm{x}}, m_\pi, 0]
+ 
\frac{2}{9}
g_{\D N}^2
I^{22}_{5/2} 
[x q \hat{\bm{x}}, m_\pi, \D]
\right\},
\end{equation}
where we have explained above how the $x$-integrals are to be performed
analytically. From this correction, we can form the ratio $\tilde{R}(L)
= \d \tilde F_2^{(I)} (0, L) / \d  F_2^{(I)} (q, L)$. This ratio has the 
same behavior as $R(L)$ shown in Figure~\ref{f:R}. The only difference
is that $\tilde{R}(L)$ is $\sim 10 \%$ larger. The careful expansion
leading to $\d \tilde F_2^{(I)} (0, L) $ thus
actually does worse to describe the finite volume matrix element
than the naive expansion, $\d F_2^{(I)} (0, L)$.   
The asymptotic approach to unity also sets in for $\tilde{R}(L)$ at 
$m_\pi L \approx 90$. 
Of course at these volumes any effect is negligible. 
To compare these approximations to the finite volume matrix element, 
we plot each individually as a function of $m_\pi L$ in Figure~\ref{f:compare}. 
For all practical purposes, beyond 
$m_\pi L \approx 5$ 
both approximations give reasonable values for the finite volume correction to the current matrix element.

\begin{figure}
\epsfig{file=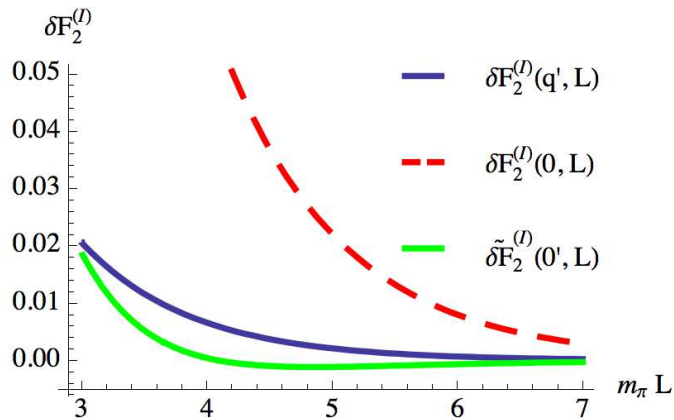,angle=270,width=9cm}
\caption{Plot of the finite volume modifications at the momentum transfer 
$\bm{q}' = \frac{2 \pi}{L} (1,1,0)$.
The finite volume form factor 
$\d  F_2^{(I)} (q', L)$, 
the naive expansion 
$\d  F_2^{(I)} (0, L) $,
and the asymptotic approximation 
$\d  \tilde{F}_2^{(I)} (0', L) $,  
are plotted  as a function of 
$m_\pi L$. 
}
\label{f:compare2}
\end{figure}

If the simplifying choice of lattice kinematics is not made, 
the correction 
$\d F_2^{(I)}(0,L)$ 
ceases to give a good estimate
for the volume corrections to the current. 
We can investigate this by choosing the momentum transfer 
to be 
$\bm{q}' = \frac{2 \pi}{L} (1, 1, 0)$ 
and returning to Eq.~\eqref{eq:Pauli} to determine the volume effect. 
This effect we denote by 
$\d F_2^{(I)} (q', L)$ 
which has the form
\begin{eqnarray}
\d F_2^{(I)} (q', L)
&\equiv&
\frac{L M_N}{2 \pi i}
\Big[ 
\d \langle N(\bm{q}') \downarrow | \hat{\bm{y}} \cdot \bm{J}_{\text{Iso}} | N (\bm{0}) \downarrow \rangle_L 
- 
\d \langle N(\bm{q}') \uparrow | \hat{\bm{y}} \cdot \bm{J}_{\text{Iso}} | N (\bm{0}) \uparrow \rangle_L 
\Big]
\notag \\
&=& - 
\frac{3 M_N}{f^2}
\Bigg\{
g_A^2 
\left[
f^{22} (q'^2, \bm{q}', 0)
-
f^{21} (q'^2, \bm{q}', 0)
\right]
+
\frac{2}{9}
g_{\D N}^2
\left[
f^{22} (q'^2, \bm{q}', \D)
-
f^{21} (q'^2, \bm{q}', \D)
\right]
\Bigg\}.
\notag \\
\label{eq:start2}
\end{eqnarray}
Notice we have scaled the matrix elements by a factor of 
$2 \pi / L$, 
not by 
$|\bm{q}'|$. 
This is because the overall momentum factor in the matrix element is 
$2 \pi / L$ 
by virtue of Eqs.~\eqref{eq:spinpol} and \eqref{eq:Pauli}. 
Expanding in large 
$ m_\pi L$ 
but fixed
$\bm{q}'^2 L^2$, 
we can derive the asymptotics of 
$\d F_2^{(I)} (q', L)$, 
which we denote by the function
$\d \tilde{F}_2^{(I)} (0', L)$ 
and is given by
\begin{equation}
\d \tilde{F}_2^{(I)} (0', L) = \d F_2^{(I)} (q', L) \Bigg|_{P_\pi(x, q'^2) = 1}
.\end{equation}
The resulting 
$x$-integrals 
in 
$\d \tilde{F}_2^{(I)} (0', L)$
involve products of two $x$-dependent Jacobi elliptic-theta functions. These 
can be evaluated analytically as shown in Appendix~\ref{FVF}. 
The notation $0'$ reflects that the asymptotic 
form still depends on the mode numbers (in this case
$n_x = n_y = 1$).
We can now compare finite volume effects in the current matrix 
element for the momentum 
$\bm{q}' = \frac{2 \pi}{L}(1,1,0)$. 
The exact correction 
$\d F_2^{(I)} (q', L)$, 
naive expansion
$\d F_2^{(I)} (0, L)$,
and asymptotic correction
$\d \tilde{F}_2^{(I)} (0', L)$
are each plotted as a function of 
$m_\pi L$ 
in Figure~\ref{f:compare2}. 
From the figure we see that naive guess 
$\d F_2^{(I)}(0,L)$ 
ceases to give a good description of the volume effect. 
The asymptotic formula
$\d \tilde{F}_2^{(I)}(0',L)$ 
approaches 
$\d F_2^{(I)}(q',L)$
for reasonably large values of 
$m_\pi L$ 
as it should. There is a crossover for 
$m_\pi L \approx 3$ 
which is why the curves start out in 
near agreement on the plot.

Now we remind the reader that 
$\d  F_2^{(I)} (0, L)$ 
was derived as the finite volume modification 
to the magnetic moment, which we denote 
$\d \mu_N^{(I)}(L)$. 
Above we have
been considering this term as a correction to the finite volume form factors. 
The question remains: if we use lattice data for the current matrix element, 
to what extent is the finite volume correction to the magnetic moment
described by 
$\d \mu_N^{(I)}(L)$? 
Based on our discussion leading up to
Eq.~\eqref{eq:est}, we expect a poor description of the volume effect
with current lattice sizes. 
We can only describe the behavior qualitatively 
because it depends on precisely how the magnetic 
moment is extracted from the lattice data.

Generally speaking what happens in any procedure
(including modeling the momentum transfer dependence)
is some type of weighting over the values of the current matrix element 
for the smallest momentum modes. 
The simplest way to do this is to take the slope of the current matrix
elements for the two smallest non-vanishing momenta. 
By Eq.~\eqref{eq:spinpol}, the slope should be the magnetic moment. 
Now consider the volume corrections to the current matrix element
at each of the sampled momenta.
For nearby modes, the volume 
effect has a good degree of cancellation and is hence reduced. 
Compare this to 
$\d F_2^{(I)}(0,L)$
which provides an otherwise decent 
estimate of either of the individual values of the current matrix element,
and consequently a poor estimate of their difference. 
We expect this argument to qualitatively generalize when one considers
an arbitrary weighting of contributions from small momentum modes. 
This constitutes a rough description of how one models the momentum transfer dependence of form factors.

The safest route in dealing with volume corrections is to deduce them 
for the momentum modes at which the current has been calculated.
Use the effective theory, if possible, 
to isolate the infinite volume values of the form factor at these data points, 
then perform a momentum extrapolation to extract the infinite volume 
magnetic moment. Carrying out the procedure in the opposite order
most likely leads to specious volume dependence.

\subsection{Form Factor Puzzle} \label{puzzle}

Finally we explore a surprising consequence of our
finite volume expressions for current matrix elements. 
We show that values of current matrix elements 
calculated at two different momenta sharing the 
same value of momentum transfer squared generally differ. 
This puzzling feature is a manifestation of the lack of rotational 
invariance in lattice data.

We consider volume corrections to the isovector Pauli 
form factor and thus return to the general expression in 
Eq.~\eqref{eq:Pauli}. Above we have determined the
effect of the finite volume at 
$\bm{q}' = \frac{2 \pi}{L} (1,1,0)$.
Now we additionally determine the effect at
the momentum
$\bm{q}'' = \frac{2 \pi}{L}(1,0,1)$. 
Of course 
$\bm{q}'^2 = \bm{q}''^2$, 
and in infinite volume the form factor must 
be the same by rotational invariance. 
For both momenta, we utilize the 
$y$-component 
of the current.%
\footnote{
A third choice of momentum, 
$\bm{q}''' = \frac{2 \pi}{L} (0,1,1)$,
additionally has the same momentum transfer squared. 
Using the $y$-component of the current in Eq.~\eqref{eq:spinpol}, 
this momentum yields zero for spin-polarized matrix elements 
in infinite volume.  Any signal from $\bm{q}'''$
on the lattice would be purely a volume effect.
}
\begin{figure}
\epsfig{file=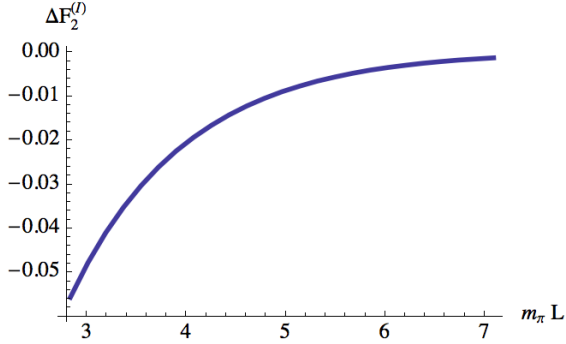,angle=270,width=9cm}
\caption{Plot of the form factor difference 
$\D F_2^{(I)} = F_2^{(I)}(\bm{q}'^2) - F_2^{(I)}(\bm{q}''^2)$,
with 
$ \bm{q}'^2 = \bm{q}''^2 = 8 \pi^2 / L^2$,   
as a function of 
$m_\pi L$. 
Rotational invariance at infinite volume requires 
this difference to vanish. 
}
\label{f:difference}
\end{figure}
To compare the value of the form factor at these
differing momenta, we form the difference $\D F_2^{(I)}$ given by
\begin{eqnarray}
\D F_2^{(I)} 
&=&
\frac{L M_N}{2 \pi i}
\Big[ 
\langle N(\bm{q}') \downarrow | \hat{\bm{y}} \cdot \bm{J}_{\text{Iso}} | N (\bm{0}) \downarrow \rangle_L 
- 
\langle N(\bm{q}') \uparrow | \hat{\bm{y}} \cdot \bm{J}_{\text{Iso}} | N (\bm{0}) \uparrow \rangle_L 
\Big]
\notag \\
&& 
- 
\frac{L M_N}{2 \pi i}
\Big[ 
\langle N(\bm{q}'') \downarrow | \hat{\bm{y}} \cdot \bm{J}_{\text{Iso}} | N (\bm{0}) \downarrow \rangle_L 
- 
\langle N(\bm{q}'') \uparrow | \hat{\bm{y}} \cdot \bm{J}_{\text{Iso}} | N (\bm{0}) \uparrow \rangle_L 
\Big]
.\end{eqnarray}
Accordingly the infinite volume pieces of the current matrix elements cancel out of 
$\D F_2^{(I)}$.  Thus 
$\D F_2^{(I)}$ 
gives a measure of how much calculations of
the form factor at 
$\bm{q}'^2 = \frac{8 \pi^2}{L^2}$ 
and 
$\bm{q}''^2 = \frac{8 \pi^2}{L^2}$ 
differ. 
Using the expression for the volume effect, Eq.~\eqref{eq:Pauli},
as well as expressions from Appendix~\ref{FVF}, 
we determine 
$\D F_2^{(I)}$ 
to be
\begin{equation}
\D F_2^{(I)}
=
- 
\frac{6 M_N}{f^2}
\left[ 
g_A^2  f ( 0 ) + \frac{2}{9} g_{\D N}^2  f(\D)
\right]
,\end{equation}
where the function $f(\D)$ is given by
\begin{eqnarray}
f(\D) 
&=& 
\frac{1}{24 \pi^{3/2} L}
\int_0^1 dx \int_0^\infty d \tau
\sqrt{\tau}
e^{- (m_\pi^2 P_\pi(x, q'^2)^2 - \D^2) L^2 / 4 \tau}
\Erfc \left( \frac{\D L }{2 \sqrt{\tau}} \right)
\notag \\
&&
\times
\left[
\vartheta''_3( \pi x, e^{- \tau})
\vartheta_3( \pi x, e^{-\tau})
\vartheta_3(0,e^{-\tau})
-
\vartheta'_3( \pi x, e^{- \tau})^2
\vartheta_3(0,e^{-\tau})
-
\vartheta''_3( 0, e^{- \tau})
\vartheta_3( \pi x, e^{-\tau})^2
\right].
\notag\\
\label{eq:mixedup}
\end{eqnarray}
We plot $\D F_2^{(I)}$  in Figure~\ref{f:difference}, and indeed find that the finite volume
form factor does not respect rotational invariance. 
From the Figure, moreover, we see that for asymptotically large volumes
rotational invariance is restored. 
This can easily be demonstrated analytically 
from Eq.~\eqref{eq:mixedup}.
This is a general feature of finite volume form factors.
One can easily 
verify the lack of rotational invariance 
for the isoscalar combination as well as charge form factor, for example.

The fact that different momenta (which share the same magnitude)
yield differing current matrix elements is already evident from the general 
structure of the finite volume effective action, Eq.~\eqref{eq:EFTL} 
[as well as Eq.~\eqref{eq:EFTLL} in Appendix~\ref{B}].
The new couplings allowed on a torus break rotational invariance.
While these couplings vanish at one-loop order in heavy baryon \CPT, 
there are indeed non-vanishing couplings at higher orders in the 
derivative expansion.
These terms are not suppressed, however, because all terms in the
derivative expansion in powers of 
$\bm{q}^2 L^2$ 
are order one.

In confronting actual lattice data for current matrix elements, 
we cannot attribute all differences at identical momentum 
transfer squared to volume effects. 
The discretization
also breaks rotational invariance. 
Thus at infinite 
volume but finite lattice spacing,  matrix elements
are only hypercubic invariant functions of the momentum.
Generally differences seen in lattice form factor data
arise from the violation of rotational invariance from 
both volume and discretization effects.  
Physically one expects the dominant source of 
difference to be from the volume (discretization) at
small (large) momentum transfer.

\section{Summary} \label{summy}

Above we have analyzed the finite volume current 
matrix elements of the nucleon. Focusing specifically
on the isovector contribution, we have deduced in 
Eq.~\eqref{eq:Pauli}
the finite volume correction to lattice calculations 
of the electromagnetic three-point function. 
This finite volume result does not have an interpretation 
as a correction to the Pauli form factor. A decomposition
of matrix elements in terms of Dirac and Pauli form factors makes use of 
Lorentz and gauge invariance. The former is not relevant 
for a torus, while the latter has new features in finite volume 
due to the special gauge transformation of the zero-mode.

For the smallest available momentum transfer on the lattice,
$q = 2 \pi / L$,  
we have shown numerically that one cannot
series expand finite volume results in powers of the 
photon momentum.  This result applies generally, not just 
to current matrix elements. Consider a simple Feynman 
diagram with momentum insertion in finite volume. 
Using a Feynman parameter to combine denominators, 
the typical contribution has the form
\begin{equation}
\propto 
\int_0^1 dx 
\sum_{\bm{n}} 
\left[ 4 \pi^2 
(\bm{n} + x \bm{m})^2  
+ 
(m_\pi L)^2
\left( 
1 - x (1 - x)
\frac{q^2}{m_\pi^2} 
\right)
\right]^{-\b}
,\end{equation}
where 
$\bm{q} = 2 \pi \bm{m} / L$ 
is the inserted momentum. 
In infinite volume, multipole expansions
are well defined in terms of the expansion parameter 
$\l = q^2 / m_\pi^2$.
In finite volume, this expansion remains
well behaved,  
$\l \propto (m_\pi L)^{-2}$ 
for large 
$L$. 
There are additional finite volume terms, however, that depend on the 
combination 
$\bm{q}^2 L^2 = 4 \pi^2 \bm{m}^2$. 
Such dependence cannot be expanded out even in large volumes.
Thus working at only the smallest momentum transfer, we cannot
deduce volume corrections to multipole moments, 
\emph{cf} Figure~\ref{f:R}. The approximate agreement shown in 
the figure results from simplifying assumptions about the lattice 
kinematics used to measure the current. The agreement worsens
when such choices are not made, see Figure~\ref{f:compare2}.

On the other hand, for large enough momenta the relative 
spacing between modes becomes small enough to approximate
a derivative.  In this regime, one can series expand finite volume
modifications as has been done in the literature. 
The resulting frequency expansion is not a multipole expansion, 
however, but is described by a theory of Wilson lines, 
as in Eq.~\eqref{eq:EFTL}. 
New couplings are allowed and lead to current screening, for example. 
For the isovector magnetic moment of the nucleon, 
we showed that a new zero-mode coupling allowed by symmetry 
vanishes at next-to-leading order in heavy baryon chiral perturbation theory.
Consequently we were able to recover previously derived results~\cite{Beane:2004tw}. 
In order for such results to be valid, however, one requires 
momenta \emph{small} enough for the efficacy of a low-energy effective 
theory, and at the same time \emph{large} enough in order to treat
the momentum as continuous. Satisfying such restrictions is considerably 
beyond the reach of current computing power. 
In practice, one must retain the momentum transfer dependence
of finite volume matrix elements to account for the volume effect
in lattice correlation functions.
Finally we showed that such finite volume corrections to form factors
can lead to surprising effects. Specifically current matrix 
elements evaluated at identical
$\bm{q}^2$, 
but differing 
$\bm{q}$, 
need not be the same.

\begin{acknowledgments}
We thank T.~Cohen, M.~Savage, and A.~Walker-Loud for 
discussion and/or comments.
This work is supported in part by the U.S.~Dept.~of Energy,
Grant No.~DE-FG02-93ER-40762.
\end{acknowledgments}

\appendix

\section{Glossary of Finite Volume Functions} \label{FVF}

Above we have determined the finite volume modifications
to single nucleon current matrix elements. 
In this Appendix, we give explicit formulae 
for the finite volume difference functions used in the main text. 
We use similar notation for these
functions as~\cite{Sachrajda:2004mi,Tiburzi:2006px}, 
where further discussion can be found.

In evaluating a Feynman diagram in finite volume, 
the loops contain a sum over the allowed
Fourier modes in a periodic box. The difference of this 
sum and the infinite volume integral is the finite volume effect.
As is customary,
we treat the length of the time direction as infinite.
In a heavy fermion formulation, all finite 
volume differences with momentum insertion 
can be cast in terms of the function 
$I_\beta^{i_1 \cdots i_j} (\bm{\theta},m, \d)$, 
defined by
\begin{eqnarray}
I^{i_1 \cdots i_j}_\b (\bm{\theta}, m, \d)
&=&
\int_0^\infty d\l
\left\{
\frac{1}{L^3} 
\sum_{\bm{n}} 
\frac{q^{i_1} \cdots q^{i_j}}{[(\bm{q} + \bm{\theta})^2 + m^2 + 2 \d \l + \l^2 ]^\b}
\right.
\notag \\
&& 
\phantom{spaceforspace}
\left.
- \int \frac{d \bm{q}}{(2 \pi)^3}
\frac{q^{i_1} \cdots q^{i_j}}{[(\bm{q} + \bm{\theta})^2 + m^2 + 2 \d \l + \l^2 ]^\b}
\right\}
,\end{eqnarray}
where 
$\bm{n}$
 sums over triplets of integers,
and the momentum modes satisfy the periodic 
box quantization condition,
$\bm{q} = 2 \pi \bm{n} / L$.
While general expressions for the exponentially convergent form of 
$I_\beta^{i_1 \cdots i_j} (\bm{\theta},m, \d)$
exist, we merely cite the required cases for our work. These are
\begin{eqnarray}
I_\b (\bm{\theta}, m,\d)
&=&
\frac{(L / 2)^{2 \b}}{\pi L^4   \Gamma(\beta)}
\int_0^\infty d \tau \,
\tau^{1 - \b} 
e^{- (m^2 - \d^2) L^2 / 4 \tau} 
\Erfc
\left( 
\frac{\d L}{2 \sqrt{\tau}}
\right)
\left[
\prod_{j=1}^3
\vartheta_3( \theta_j L / 2 , e^{-\tau}) - 1
\right],
\notag \\
\\
I_\b^{i_1} (\bm{\theta}, m,\d)
&=&
-\frac{1}{2 (\b - 1)}
\frac{d}{d\theta^{i_1}}
I_{\b-1}(\bm{\theta},m,\d)
- 
\theta^{i_1} I_\b
(\bm{\theta}, m,\d), \, \text{and}
\\
I_\b^{i_1 i_2} (\bm{\theta}, m,\d)
&=&
\frac{1}{4 (\b - 2)(\b - 1)}
\frac{d^2}{d\theta^{i_1} d\theta^{i_2}}
I_{\b-2}(\bm{\theta},m,\d)
+
\frac{1}{2 (\b - 1)}
\d^{i_1 i_2} I_{\b - 1} (\bm{\theta},m,\d)
\notag \\
&&
+
\frac{1}{2 (\b - 1)}
\left[
\theta^{i_1} 
\frac{d}{d\th^{i_2}}
I_{\b-1}(\bm{\theta}, m,\d)
+
\theta^{i_2} 
\frac{d}{d\th^{i_1}}
I_{\b-1}(\bm{\theta}, m,\d)
\right]
+ 
\theta^{i_1} \theta^{i_2} I_\b(\bm{\theta}, m,\d),
\notag \\
\end{eqnarray}
where $\vartheta_3 (z, q)$ is the Jacobi elliptic-theta
function of the third kind, and $\Erfc (z)$ is the complement of 
the standard error function. The finite volume physics
arising from the mass splitting parameter $\d$
is well explicated in~\cite{Arndt:2004bg}.

In expanding matrix elements about zero external momentum, 
one also has use for the functions $J_\b(m,\d)$, and $K_\b(m,\d)$
given by
\begin{eqnarray}
J_\b (m, \d) 
&=&
- \frac{1}{4 (\b -1)(\b-2)}
\frac{\partial^2}{\partial \d^2}
I_{\b - 2} (\bm{0}, m, \d)
+
\left(
1 + \frac{\d}{\b -1} \frac{\partial}{\partial \d}
\right)
I_{\b - 1} (\bm{0}, m, \d)
-
m^2 
I_{\b} (\bm{0}, m, \d),
\notag \\
\\
K_\b (m, \d)
&=&
- \frac{1}{4 (\b -1)(\b-2)}
\frac{\partial^2}{\partial \d^2}
J_{\b - 2} (\bm{0}, m, \d)
+
\left(
1 + \frac{\d}{\b -1} \frac{\partial}{\partial \d}
\right)
J_{\b - 1} (\bm{0}, m, \d)
-
m^2 
J_{\b} (\bm{0}, m, \d).
\notag \\
\end{eqnarray}
Lastly of use are Feynman parameter integrals of Jacobi elliptic-theta functions. 
\begin{eqnarray}
&& \int_0^1 dx  \,  \vartheta_3( \pi x, q) 
=
1 
,\\
&& \int_0^1 dx \,  \vartheta_3( \pi x, q)^2
=
 \vartheta_3( 0, q^2) 
,\\
&& \int_0^1 dx \,  \vartheta'_3( \pi x, q)^2
=
- \vartheta''_3( 0, q^2) 
,\\
&& \int_0^1 dx \,  \vartheta''_3( \pi x, q)  \vartheta_3( \pi x, q) 
=
\vartheta''_3( 0, q^2) 
.\end{eqnarray}
Here the primes denote derivatives with respect to the first argument.
Double primes can be traded in for single derivatives with respect to the 
second argument because the 
$\vartheta_3(z,q)$ 
satisfy a diffusion equation.

\section{Magnetic Radius} \label{B}

In this Appendix, we present the extension of the finite volume effective
action in Eq.~\eqref{eq:EFTL} to higher order in the derivative expansion. 
This will allow us to consider the finite volume corrections to the magnetic
radius. For simplicity, we ignore the charge radius and focus just on the 
spin-dependent part of the single particle effective action.

Using the cubic symmetry of the torus, it is easy to see that 
there are no spin-dependent terms with two derivatives. 
Including all spin-dependent terms with three derivatives
and at most one insertion of 
$\cW_i^{(-)}$, 
we arrive at the nucleon effective action
\begin{eqnarray}
\cL 
&=& 
\ol N 
\Big[
C_3(L)
\bm{\sigma} 
\cdot 
\partial^2 
\bm{B} 
- 
C_4(L)
\bm{\sigma} \cdot \nabla^2  \bm{B} 
-
C_5(L)
\bm{\sigma} \cdot ( \bm{\nabla} \times \nabla^2  \bm{\cW}^{(-)} )
\notag \\
&&
-
C_6(L)
\sum_i
\sigma_i
\nabla_i \nabla_i   B_i  
-
C_7(L)
\sum_i
\varepsilon_{ijk}
\sigma_i
\nabla_i \nabla_i 
\nabla_j  \cW^{(-)}_k
\Big] N  
\label{eq:EFTLL}
.\end{eqnarray}
Here the couplings 
$C_4(L)$, $C_5(L)$, $C_6(L)$, and $C_7(L)$ 
run to zero as the volume goes to infinity. 
This reflects that the corresponding operators
are forbidden by infinite volume gauge invariance, 
and Lorentz  invariance. 
The remaining coupling
$C_3(L)$,
satisfies
\begin{equation}
C_3(L) = C_3 + \delta C_3(L)
,\end{equation}
where $C_3$ is the infinite volume coupling given by
\begin{equation}
C_3 = 6  < r_M^2 >
,\end{equation}
with $< r_M^2 >$ as the mean-square magnetic radius
which we define by $- \frac{1}{6} \frac{d}{dq^2} F_2(q^2)$.
The finite volume effect
$ 6  \, \d C_3(L)$ 
is thus the correction to the magnetic radius, however, 
there are four additional operators that also make 
contributions to spin polarized current matrix elements
in finite volume.

To determine the couplings in Eq.~\eqref{eq:EFTLL},
we turn to heavy baryon chiral perturabtion theory. 
Specifically we need the result derived above for 
spin polarized current matrix elements, Eq.~\eqref{eq:Pauli}. 
We expand this result to third order in the photon 
frequency and match onto Eq.~\eqref{eq:EFTLL}.  
This will determine the isovector part of the coupling constants.
We find
\begin{eqnarray}
\delta C_3(L) 
&=&
\frac{5}{6 f^2}
\left[
g_A^2 J_{7/2} (m_\pi , 0) + \frac{2}{9} J_{7/2}(m_\pi, \D)
\right]
,\notag \\
C_4(L) 
&=&
\frac{1}{3 f^2}
\left[
g_A^2 
\left(
7 K_{9/2}(m_\pi,0) - 5 J_{7/2} (m_\pi, 0)
\right)
+ 
\frac{2}{9}
g_{\D N}^2
\left( 
7 K_{9/2}(m_\pi,\D) - 5 J_{7/2} (m_\pi, \D)
\right)
\right]
,\notag \\
C_5(L) &=&
C_6(L) = 
C_7(L) = 0
.\end{eqnarray}
The finite volume functions $J_\b (m, \d) $ and $K_\b (m, \d) $ are defined in Appendix~\ref{FVF}.
The fact that $C_3(L)$ and $C_4(L)$ are non-vanishing to this
order means that it is not possible to define the magnetic radius
at finite volume. The remaining coefficients require further terms
in the chiral expansion to be non-vanishing.  The non-relativistic 
nature of the nucleon in heavy baryon chiral perturbation theory
is likely the culprit for the vanishing of these terms at next-to-leading 
order. In the meson sector, by contrast, derivative couplings lead to
a myriad of new finite volume terms, see~\cite{Hu:2007ts}.

\bibliography{hb}

\end{document}